\documentclass[aps,prl,twocolumn,floatfix,showpacs,11pt,reprint]{revtex4-1}

\usepackage{graphicx}
\usepackage{amsmath, amsfonts, amssymb, bm}

\usepackage{amstext}
\usepackage{mathrsfs}

\begin{document}
\title{Stochasticity effects in quantum radiation reaction}
\author{N.~Neitz}
\author{A.~Di Piazza}
\email{dipiazza@mpi-hd.mpg.de}
\affiliation{Max-Planck-Institut f\"ur Kernphysik, Saupfercheckweg 1, D-69117 Heidelberg, Germany}

\begin{abstract}
When an ultrarelativistic electron beam collides with a
sufficiently intense laser pulse, radiation-reaction
effects can strongly alter the beam dynamics.
In the realm of classical electrodynamics, radiation reaction
has a beneficial effect on the electron beam as 
it tends to reduce its energy spread. Here, we show that 
when quantum effects become important, 
radiation reaction induces the opposite effect, i.e., 
the electron beam spreads out after
interacting with the laser pulse. We identify the physical 
origin of this opposite tendency in the intrinsic stochasticity 
of photon emission, which becomes substantial in the full quantum regime. 
Our numerical simulations indicated that the
predicted effects of the stochasticity can be measured 
already with presently available lasers and electron 
accelerators.
\end{abstract}

\maketitle

A deep understanding of the dynamics of electric charges 
driven by electromagnetic fields
is one of the most fundamental problems in physics, as 
it has implications in different fields,
including accelerator, radiation and high-energy physics. Apart from
its impact on practical issues, as the construction of new 
experimental devices (e.g., quantum x-free electron lasers \cite{Bonifacio_1985}),
the investigation of the dynamics of electric charges 
(electrons, for definiteness) is also of pure theoretical 
interest, as it involves in general a coupled
dynamics of the electrons and of their own electromagnetic field.

In the realm of classical electrodynamics, 
radiation-reaction (RR) effects stem from the back
reaction on the electron dynamics of the electromagnetic
field generated by the electron itself while being
accelerated by a background electromagnetic field \cite{DiPiazza:2012rev,Rohrlich:2002}.
The Landau-Lifshitz (LL) equation has been 
recently identified as the classical equation of motion of
an electron, with mass $m$ and charge $e<0$, which includes RR effects self-consistently 
\cite{Landau:1975,Spohn:2000,Rohrlich:2002,DiPiazza:2012rev,Spohn:2004,Rohrlich_b_2007},
although alternative models have been suggested
\cite{Sokolov_2009,Hammond_2010}. The analytical
solution of the LL equation in a plane-wave field \cite{DiPiazza:2008} 
shows that if an electron impinges with initial four-momentum $p_0^{\mu}$ onto a 
plane-wave field (electric-field amplitude $E_0$, central
angular frequency $\omega_0$ and propagating along the direction $\bm{n}$),
RR effects substantially affect the electron dynamics, 
if the parameter $R_c=\alpha\chi_0\xi_0$ is of the order
of unity (see also \cite{Koga_2005}). Here, $\alpha=e^2$ 
is the fine-structure constant, $\chi_0=((np_0)/m)E_0/E_{cr}$ 
is the so-called quantum nonlinearity parameter, with $n^{\mu}=(1,\bm{n})$ 
and $E_{cr}=m^2/|e|=1.3\times 10^{16\;\text{V/cm}}$,
and $\xi_0=|e|E_0/m\omega_0$ is the classical nonlinearity 
or relativistic parameter (units with $\hbar=c=1$
are used throughout). It is worth noting that, although $\chi_0$ 
is much smaller than unity in the realm of classical electrodynamics \cite{DiPiazza:2012rev},
the parameter $R_c$ can be of the order of unity \cite{Landau:1975,Koga_2005,DiPiazza:2008}.
The parameter $R_c$ represents the average energy radiated
by the electron in one laser period in units of the initial electron energy,
and for an ultrarelativistic electron initially counterpropagating
with respect to the laser field with energy $\varepsilon$, 
it is $R_c=3.2\,\varepsilon[\text{GeV}]I_0[10^{23}\;\text{W/cm$^2$}]/\omega_0[\text{eV}]$, 
with $I_0=E_0^2/4\pi$ being the laser pulse peak intensity. 
The numerical value of the parameter $R_c$ shows the generally 
demanding requirements to observe large RR effects and 
it explains why the LL equation still lacks an experimental confirmation (see 
\cite{Koga_2005,Di_Piazza_2009,Harvey_2011,Thomas_2012} for recent experimental
proposals). The expression of the parameter $R_c$ is also in agreement
with the well-known classical result that more energetic 
particles radiate more at given other conditions \cite{Jackson_b_1975}. In turn, 
this explains physically the beneficial effect
of RR when it is included, e.g., in the investigation
of the production of electron \cite{Zhidkov_2002} and 
ion \cite{Naumova_2009,Chen_2010,Tamburini_2010,Tamburini_2011} 
bunches in laser-plasma interaction. In fact, it is 
found that RR acts as a cooling mechanism and its effects 
render the energy spectra of the produced particle bunches
more monochromatic than if RR is not included.

In this Letter we show that when quantum effects become important
RR induces exactly the opposite behavior and makes
the energy distribution of an electron beam
initially counterpropagating with respect to a strong laser field
broader as it was before the interaction.
We explain this striking difference between
classical and quantum RR relating it to the
stochastic nature of the emission of radiation, which becomes
substantial in the quantum regime, and indicating that
quantum effects amount to add a stochastic term
in the LL classical equation.
By means of numerical simulations we show
that the broadening of the electron distribution in the quantum regime, is measurable
in principle with presently available technology even
in an all-optical setup. Our results
are relevant for future laser-based electron accelerators,
indicating that one cannot rely on the beneficial effects
of RR on the energy spread of the electron beam at sufficiently
high electron energies that quantum effects become important. 
We note that the stochastic nature of photon emission has 
instead been shown to lower the laser intensity threshold
at which electromagnetic cascades are generated \cite{Duclous2011}.

Taking into account exactly RR in the full strong-field QED regime is a formidable task,
as it amounts to determine completely the $S$-matrix, describing
the interaction of the electron-positron field with
the radiation field in the presence of the strong background
electromagnetic field \cite{DiPiazza:2010mv,DiPiazza:2012rev}. 
Thus, we limit here to the so-called ``nonlinear 
moderately-quantum'' regime \cite{DiPiazza:2010mv}, 
where: 1) $\xi_0\gg 1$, such that nonlinear effects in the laser
field amplitude are large; 2) $\chi_0\lesssim 1$, such that 
nonlinear QED effects are already important, but electron-positron 
pair production is still negligible. In this regime, 
RR effects on the electron dynamics in 
a strong plane-wave field mainly 
stem from the sequential emission of many 
photons by the electron, and they can be investigated by means of
a kinetic approach \cite{Baier:1998vh,Khokonov_2004,Sokolov:2010am}
(see \cite{DiPiazza:2010mv}, for an alternative, microscopic approach).
In this approach, the electrons and the photons are described
by distribution functions in phase space, which obey to
``kinetic'' equations. Since electron-positron
pair production is neglected: 1) the distribution function of positrons
can be assumed to vanish identically; 2) the kinetic equation
for the electron distribution function is not coupled
to that of the photons \cite{Baier:1998vh,Khokonov_2004,Sokolov:2010am}.
Another realistic approximation, which allows us to avoid technical complications
in favor of a clearer physical understanding, is to consider
an electron bunch initially counterpropagating with respect
to the laser field and with a typical energy $\varepsilon^*\gg m\xi_0$.
This is the case, for example, in the realistic situation
of a electron bunch with typical energy 
$\varepsilon^*=1\;\text{GeV}$ colliding head-on
with an optical ($\omega_0=1.55\;\text{eV}$) laser field of 
intensity $10^{22}\;\text{W/cm$^2$}$ \cite{Yanovsky_2008} for which $m\xi_0=25\;\text{MeV}$. 
The condition $\varepsilon^*\gg m\xi_0$ ensures that the
transverse momentum of the electrons (with respect to the initial propagation
direction) remains much smaller than the longitudinal
one in passing through the plane wave \cite{Landau_b_2_1975},
and this reduces the present problem to a one-dimensional one.

By assuming that the plane wave propagates along the positive
$y$ direction and that it is linearly-polarized along the $z$ direction,
we can write its electric field as $\bm{E}(\varphi)=E_0f(\varphi)\hat{\bm{z}}$,
where $\varphi=\omega_0(t-y)$ is the laser phase and $f(\varphi)$ 
is the pulse-shape function such that $|f(\varphi)|_{\text{max}}=1$. 
If $p^{\mu}=(\varepsilon,\bm{p})$ is the
four-momentum of an electron, it is convenient to introduce
the quantity $p_-=\varepsilon-p_y$, which is a constant of motion
in the plane-wave field under consideration \cite{Landau_b_2_1975}. However, if the
electron emits a photon with four-momentum $k^{\mu}=(\omega,\bm{k})$,
then its four-momentum changes to $p^{\prime\mu}=(\varepsilon',\bm{p}')$
and $p'_-=p_--k_-$, with $p'_-=\varepsilon'-p'_y$ and $k_-=\omega-k_y$.
The single-photon emission probability per unit phase $\varphi$ 
and per unit $u=k_-/(p_--k_-)$ in the ultrarelativistic regime 
$\xi_0\gg 1$ reads \cite{Ritus:1985}
\begin{equation}
\label{ToniProb}
\begin{split}
\frac{dP_{p_-}}{d\varphi du}=&\frac{\alpha}{\sqrt{3}\pi}\frac{m^2}{\omega_0p_-}\frac{1}{(1+u)^2}\left[\left(1+u+\frac{1}{1+u}  \right)
\right.\\
&\left.\times\text{K}_{\frac{2}{3}}\left( \frac{2u}{3\chi(\varphi,p_-)}\right)-\int_\frac{2u}{3\chi(\varphi,p_-)}^\infty dx\, \text{K}_{\frac{1}{3}}(x) \right],
\end{split}
\end{equation}
where $\text{K}_\nu(\cdot)$ is the modified Bessel function of $\nu$th 
order and where $\chi(\varphi,p_-)=(p_-/m)|E(\varphi)|/E_{\text{cr}}$, 
with $E(\varphi)=E_0f(\varphi)$. Since the probability in Eq. (\ref{ToniProb})
depends only on the phase-space variables $\varphi$ and $p_-$, it is possible
to describe the electron beam via an electron distribution $n_e(\varphi,p_-)$,
which satisfies the kinetic equation (see Ref. \cite{Baier:1998vh})
\begin{equation}
\label{Kinetic}
\frac{\partial n_e}{\partial\varphi}=\int_{p_-}^\infty dp'_-\,\frac{dP_{p'_-}}{d\varphi dp_-}n'_e-\int_0^{p_-}dk_-\frac{dP_{p_{-}}}{d\varphi dk_-}\,n_e
\end{equation}
with $n_e=n_e(\varphi,p_-)$, $n'_e=n_e(\varphi,p'_-)$ and
\begin{align}
\frac{dP_{p'_-}}{d\varphi dp_-}&=\frac{p'_-}{p^2_-} \left.\frac{dP_{p'_-}}{d\varphi du}\right\vert_{u=\frac{p'_--p_-}{p_-}},\\
\frac{dP_{p_-}}{d\varphi dk_-}&=\frac{p_-}{(p_--k_-)^2} \left.\frac{dP_{p_-}}{d\varphi du}\right\vert_{u=\frac{k_-}{p_--k_-}}.
\end{align}
The kinetic equation (\ref{Kinetic}) is an integro-differential equation, i.e.,
it is non-local in the momentum $p_-$. This occurrence is intimately connected
to the quantum nature of the emission of radiation. In fact, 
the emission of radiation is described quantum mechanically as the emission 
of photons, which carry energy and momentum, such that, if an electron emits a photon with momentum $k_-$, its
initial state with a given momentum $p_{0,-}$ will be coupled
to that with momentum $p_{0,-}-k_-$, with $k_-$ ranging from 0
to $p_{0,-}$. 

In order to investigate the classical limit 
of Eq. (\ref{Kinetic}) for $\chi(\varphi,p_-)\ll 1$, it is convenient to perform the change of variable
$v=(p'_--p_-)/p_-\chi(\varphi,p_-)$ ($v=k_-/(p_--k_-)\chi(\varphi,p_-)$) 
in the first (second) integral in Eq. (\ref{Kinetic}). By expanding the resulting
equation in $\chi(\varphi,p_-)$ and by keeping terms
up to the order $\chi^3(\varphi,p_-)$, one obtains the Fokker-Planck-like equation \cite{Gardiner_b_2009} (see also \cite{Landau_b_10_1981,Sokolov:2010am})
\begin{equation}
\label{FP}
\frac{\partial n_e}{\partial\varphi}=-\frac{\partial}{\partial p_-}[A(\varphi,p_-)n_e]
+\frac{1}{2}\frac{\partial^2}{\partial p_-^2}\left[B(\varphi,p_-)n_e\right]
\end{equation}
with a ``drift'' coefficient $A(\varphi,p_-)$ and a ``diffusion'' coefficient $B(\varphi,p_-)$ given by
\begin{align}
\label{A}
A(\varphi,p_-)&=-\frac{2\alpha m^2}{3\omega_0}\chi^2(\varphi,p_-)\bigg[1-\frac{55\sqrt{3}}{16}\chi(\varphi,p_-)\bigg],\\
\label{B}
B(\varphi,p_-)&=\frac{\alpha m^2}{3\omega_0}\frac{55}{8\sqrt{3}}p_-\chi^3(\varphi,p_-),
\end{align}
respectively. It is worth observing that this equation is no longer an integro-differential
equation but rather a partial differential equation. In other words, when quantum photon-recoil effects
become smaller and smaller, the distribution function of electrons with
a momentum $p_-$ depends only on its values close to $p_-$ and
its dynamics is local. On this respect, we also note that higher-order corrections
in $\chi(\varphi,p_-)$ would result in the appearance of
terms proportional to higher and higher derivatives of $n_e(\varphi,p_-)$
with respect to $p_-$.

If we first consider only the terms 
proportional to $\chi^2(\varphi,p_-)$ in Eq. (\ref{FP}), the latter equation
has the form of a Liouville equation:
\begin{equation}
\label{Kinetic_cl}
\frac{\partial n_e}{\partial\varphi}=-\frac{\partial}{\partial p_-}\left(n_e\frac{d p_-}{d\varphi}\right),
\end{equation}
where
\begin{equation}
\label{pminLL}
\frac{d p_-}{d\varphi}=-\frac{I_{cl}(\varphi,p_-)}{\omega_0},
\end{equation}
with $I_{cl}(\varphi,p_-)=(2/3)\alpha m^2 \chi^2(\varphi,p_-)$
being the classical intensity of radiation \cite{Landau_b_2_1975}. 
Equation (\ref{pminLL}) is exactly the classical single-particle equation 
for the momentum $p_-$ resulting from the 
LL equation \cite{DiPiazza:2008} (see also \cite{Elkina:2010up}). In other words, the terms
in Eq. (\ref{FP}) proportional to $\chi^2(\varphi,p_-)$ 
describe the classical dynamics of the electron distribution 
including RR. The fact that Eq. (\ref{Kinetic_cl}) has the
form of a Liouville equation implies, as it must be,
that the classical dynamics of the electron distribution
is deterministic \cite{Gardiner_b_2009}. Also, since
the single-particle equation (\ref{pminLL}) admits
the analytical solution \cite{DiPiazza:2008}, $p_-(\varphi;p_{0,-})=p_{0,-}/h(\varphi,p_{0,-})$ with
\begin{equation}
\label{LLansol}
h(\varphi,p_{0,-})=1+\frac{2}{3}\alpha\frac{p_{0,-}}{\omega_0}\frac{E_0^2}{E^2_{\text{cr}}}\int_0^\varphi d\varphi' f^2(\varphi')
\end{equation}
for an electron with initial momentum $p^{\mu}(0)=p^{\mu}_0=
(\epsilon_0,\bm{p}_0)$ ($p_{0,-}=\epsilon_0-p_{0,y}$), one can write
explicitly the exact analytical solution of Eq. (\ref{Kinetic_cl})
by means of the method of characteristics. If the distribution 
$n_e(0,p_-)$ at the initial phase $\varphi=0$ is given, for example, 
by the Gaussian distribution $n_e(0,p_-)= N\exp[-(p_--p_-^\ast)^2/2\sigma^2_{p_-}]$,
where $N$ is a normalization factor, $p_-^\ast$ is the average value of 
$p_-$ and $\sigma_{p_-}$ is the standard deviation \cite{Footnote}, 
then the solution of Eq. (\ref{Kinetic_cl}) reads
\begin{equation}
\label{ansoldistr}
n_e(\varphi,p_-)=\frac{N}{g^2(\varphi,p_-)}\exp\left\{-\frac{1}{2\sigma^2_{p_-}}\left[\frac{p_-}{g(\varphi,p_-)}-p_-^\ast\right]^2\right\},
\end{equation}
with $g(\varphi,p_-)=h(\varphi,-p_-)$. Since $p_{0,-}$ in Eq. (\ref{LLansol}) is positive for finite values 
of $p_{0,y}$ and $p_{0,-}\to 0$ only at $p_y\to +\infty$, the function 
$g(\varphi,p_-)$ must be non-negative for all values of $\varphi$, and 
the equation $g(\varphi,p_{-,\text{max}})=0$ fixes the maximum value 
$p_{-,\text{max}}=p_{-,\text{max}}(\varphi)$ allowed for the variable 
$p_-$ at each $\varphi$. Before passing to investigate the quantum corrections
in Eq. (\ref{FP}), we observe that the classical solution in Eq. (\ref{LLansol})
is such that $0<\partial p_-(\varphi;p_{0,-})/\partial p_{0,-}<1$ for $\varphi>0$
and this ensures that, due to RR effects, the difference $\Delta p_-(\varphi)$ 
between the momenta of two electrons decreases for increasing 
values of $\varphi$. This implies that RR effects tend to decrease the energy width
of the electron distribution in agreement with previous results 
\cite{Tamburini_2010,Tamburini_2011}. Also, if $\sigma_{p_-}\ll p^*_-$ in Eq. (\ref{ansoldistr}),
it can be seen that the distribution $n_e(\varphi,p_-)$ is approximately
a Gaussian with effective width $\sigma_{p_-}(\varphi)\approx \sigma_{p_-}/h^2(\varphi,p^*_-)$
decreasing at increasing $\varphi$'s.

The quantum corrections in Eq. (\ref{FP}) to the classical kinetic equation 
(\ref{Kinetic_cl}) stem from two different contributions. 
The first one affects the drift coefficient $A(\varphi,p_-)$ (see Eq. (\ref{A}))
and it corresponds to the leading quantum correction to the total intensity of radiation found 
in \cite{Baier:1998vh,Ritus:1985}. This correction, does not change the
structure of the classical kinetic equation (\ref{Kinetic_cl}) but
only the ``effective'' momentum change per unit phase.
Since this leading quantum correction is negative, we expect
that it tends to decrease the reduction of the
width with respect to the classical prediction. However, 
by replacing the classical intensity of
radiation $I_{cl}(\varphi,p_-)$ with the corresponding
quantum one $I_q(\varphi,p_-)$ (see, e.g., Eq. (83) on pg. 522 in \cite{Ritus:1985}),
the resulting Liouville equation would still predict
a reduction of the width of the electron distribution function.
Although this statement can be proven mathematically, 
it can be intuitively understood as a physical consequence
of the fact that more energetic electrons \emph{on average} 
emit more radiation. On the other hand, however, 
the second leading quantum correction
corresponds to the diffusion coefficient $B(\varphi,p_-)$
in Eq. (\ref{B}) and it alters the structure of 
the classical kinetic equation. The appearance of
a diffusion-like term in the kinetic equation
of the electron distribution is intimately connected
to the stochastic nature of the quantum emission
of photons. According to the theory of stochastic
differential equations, in fact, the Fokker-Planck-like
equation (\ref{FP}) is related to the single-particle 
stochastic equation $dp_-=-A(\varphi,p_-)d\varphi+\sqrt{B(\varphi,p_-)}dW$,
where $dW$ represents an infinitesimal stochastic function \cite{Gardiner_b_2009}. 
The diffusion term in Eq. (\ref{FP}) is responsible of the broadening
of the distribution function. In the case, for example,
of a Gaussian distribution function assumed to be well peaked at $\varphi$ around the classical
value $p_-(\varphi;p^*_-)$ (see Eq. (\ref{LLansol})), it can easily
be shown, that if $\sigma_{p_-}$ is its initial width, then
\begin{equation}
\label{sigma}
\sigma_{p_-}(\varphi)\approx\frac{1}{h^2(\varphi,p^*_-)}\left[\sigma^2_{p_-}+\int_0^{\varphi}d\varphi'B(\varphi',p^*_-)\right]^{1/2}.
\end{equation}
This result clearly shows the opposite influence of the classical drift term and of the
quantum diffusion term on the width of the electron distribution. It is worth noting
that the correction to the width arising from the quantum corrections
in the drift coefficient $A(\varphi,p)$ is found to be smaller than
the correction proportional to the diffusion term by a factor $\sigma_p^2/p_-^{*,2}\ll 1$,
and it has been neglected in the approximated expression (\ref{sigma}).
We warn the reader about the fact that a formal solution of the Fokker-Planck
equation, for example, in the case of an initial $\delta$-like momentum
distribution and vanishing drift term, would predict, due to the spreading
in the momentum distribution, 
the spurious presence of particles with
momentum larger than the initial one. This indicates that, for a
completely consistent treatment, the full equation (\ref{Kinetic}) has
to be employed, which will be carried out below numerically.

The above effect on the broadening of the electron momentum distribution
can also be interpreted in terms of the entropy 
$S(\varphi)=-\int_0^\infty dp_-\, n_e(\varphi,p_-)\ln[n_e(\varphi,p_-)/n_0]$ 
associated to the the electron distribution, where the Boltzmann constant 
has been set equal to unity and where the physically ineffective constant 
$n_0$ can be chosen, for example, such that $S(0)=0$. By employing this definition 
and Eqs. (\ref{FP}), (\ref{A}) and (\ref{B}), it results
\begin{equation}
\label{EntropyChi3}
\begin{split}
\frac{dS}{d\varphi}=&-\frac{4\alpha m^2}{3\omega_0}\int_0^\infty \frac{dp_-}{p_-}\chi^2(\varphi,p_-)n_e\Bigg\{ 1\\
&\left.-\frac{55\sqrt{3}}{32}\chi(\varphi,p_-)\left[1+\frac{1}{6}\frac{p_-^2}{n_e^2}\left(\frac{\partial n_e}{\partial p_-}\right)^2\right]\right\}.
\end{split}
\end{equation}
This result further corroborates the idea that, while the classical
``deterministic'' evolution of the electrons
implies a reduction of the entropy of the electrons,
quantum corrections tend to increase it.

In order to show that the effects discussed above
can be in principle measured with presently available
laser and electron accelerator technology, we consider
below two numerical examples. In both cases 
we assume a laser pulse with $f(\varphi)=\sin^2(\varphi/2N_L)\sin(\varphi)$
for $\varphi\in [0,\varphi_f]=[0,2 N_L\pi]$ and zero elsewhere,
where $N_L$ is the number of laser cycles and with $\omega_0=1.55\;\text{eV}$,
and an initial Gaussian electron distribution with a total number of 1000 electrons.

In the first numerical example, we choose the laser and electron
parameters such that quantum effects are negligible,
whereas RR effects are relatively large. We set 
$I_0=4.3\times 10^{20}\;\text{W/cm$^2$}$, $p_-^*=84\;\text{MeV}$ 
(note that $\varepsilon^*\approx p_-^*/2=42\;\text{MeV}$) such
that $\chi^*=(p^*_-/m)(E_0/E_{cr})\approx 5\times 10^{-3}$, $\sigma_{p_-}=8.4\;\text{MeV}$, and $N_L=1600$, 
corresponding to a pulse duration of about $4\;\text{ps}$.
The results for the initial and final distribution are shown in Fig. 1.
\begin{figure}
\includegraphics[width=\linewidth]{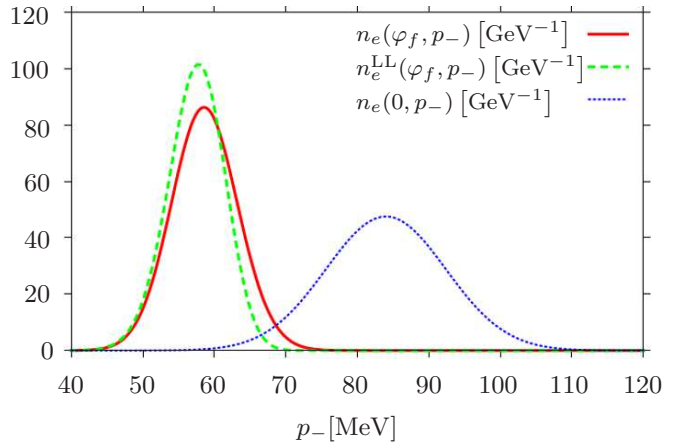} 
\caption{Comparison of the initial electron distribution (dotted, blue line) and the final electron distribution according to Eq. (\ref{Kinetic}) (solid, red line) and to Eq. (\ref{ansoldistr}) (dashed, green line). The laser and the electron distribution parameters are given in the text.}
\end{figure}
As expected, the final distribution $n_e(\varphi_f,p_-)$, calculated by solving numerically
Eq. (\ref{Kinetic}) (solid, red line) and the classical analytical solution $n_e^{\text{LL}}(\varphi_f,p_-)$ (see Eq. (\ref{ansoldistr})) are very similar and both show a reduction of the width from
the initial value $8.4\;\text{MeV}$ to the final one $4.7\;\text{MeV}$. 
In the second numerical example, instead, we want to
probe the quantum regime and we set $I_0=2\times 10^{22}\;\text{W/cm$^2$}$ \cite{Yanovsky_2008}, 
$p_-^*=2\;\text{GeV}$ ($\varepsilon^*\approx 1\;\text{GeV}$) and $\sigma_{p_-}=0.2\;\text{GeV}$ 
corresponding to $\chi^*=0.8$, and $N_L=10$ corresponding to about $30\;\text{fs}$. Electron beams
with such energies are nowadays available not only
in conventional accelerators but also by
employing plasma-based electron accelerators \cite{Leemans_2006} (see also \cite{Mangles_2004}),
allowing in principle for an all-optical setup.
The results of our numerical simulations are shown
in Fig. 2.
\begin{figure}
\includegraphics[width=\linewidth]{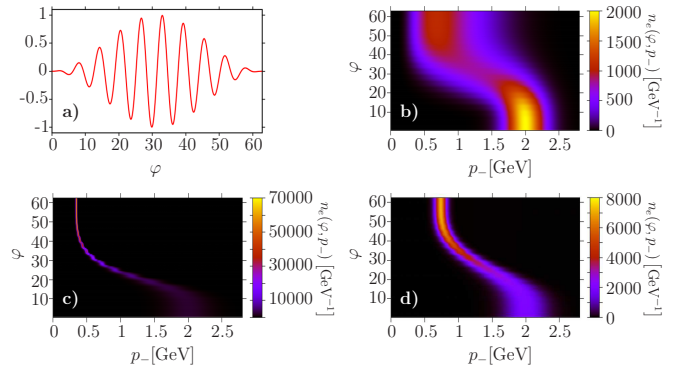}
\caption{Time evolution of the electron distribution for a 10-cycle $\sin^2$-like laser pulse (part a)) according to Eq. (\ref{Kinetic}) (part b)), to Eq. (\ref{ansoldistr}) (part c)) and to Eq. (\ref{Kinetic_cl}) and Eq. (\ref{pminLL}) with the replacement $I_{cl}(\varphi,p_-)\rightarrow I_q(\varphi,p_-)$ (part d)). The laser and the electron distribution parameters are given in the text.}
\end{figure}
The figure shows that in the quantum regime the full quantum
calculations based on Eq. (\ref{Kinetic}) predict a broadening of the electron
distribution (Fig. 2b), according to our analysis above. Whereas, the classical
calculations based on the exact solution in Eq. (\ref{ansoldistr})
(see Fig. 2c) predict a strong narrowing of the distribution. It is interesting to note
that, according to the discussion above Eq. (\ref{sigma}), if
we consider the classical equation (\ref{Kinetic_cl}) and
we substitute the classical intensity of radiation $I_{cl}(\varphi,p_-)$
with the quantum intensity $I_q(\varphi,p_-)$ (see, e.g., 
\cite{Baier:1998vh,Ritus:1985}), the corresponding results (see Fig. 2d)
still predict a narrowing of the distribution function.
This clearly supports the idea that the broadening of the
electron distribution is an effect of the importance of the stochasticity
of the emission of radiation, which becomes substantial 
in the quantum regime.

In conclusion, we have demonstrated that the importance of the stochastic
nature of the emission of radiation in the quantum regime, 
has a profound impact on the evolution
of an electron beam passing through an intense laser
field. The stochasticity, in fact, induces a broadening
on the electron momentum distribution, whereas classical
theory of RR even predicts a narrowing of the
distribution itself. A numerical example has shown the
feasibility of measuring such effects by employing
already demonstrated laser intensities and electron-beam energies.

The authors would like to gratefully acknowledge useful discussions with John Kirk 
and Tom Blackburn.


\begin{thebibliography}{99}
\bibitem{Footnote} This were strictly true, if the variable $p_-$ would run from $-\infty$ to $+\infty$. However, in our case, $p_->0$, and $p_-^*$ and $\sigma_{p_-}$ are a good approximation of the average value and of the standard deviation of the distribution, respectively, if $\sigma_{p_-}\ll p_-^*$, which we assume to be fulfilled throughout the Letter.
\end{thebibliography}
\end{document}